\documentclass[aps,prl,twocolumn,groupedaddress,showpacs]{revtex4}

\usepackage{graphicx}
\usepackage{amsmath}
\usepackage{amssymb}
\begin{document}
\title{Momentum spectroscopy of 1D phase fluctuations in Bose-Einstein condensates}
\author{S.\ Richard}
\author{F.\ Gerbier}
\author{J.\ H.\ Thywissen}
\author{M.\ Hugbart}
\author{P.\ Bouyer}
\email[email: ]{philippe.bouyer@iota.u-psud.fr}
\author{A. Aspect}
%\thanks{}
\affiliation{Laboratoire Charles Fabry de l'Institut d'Optique
\footnote{UMR 8501 du CNRS}, 91403 Orsay Cedex, France}
%\homepage[]{http://atomoptic.iota.u-psud.fr/}
%
\date{\today}
\begin{abstract}
We measure the axial momentum distribution of Bose-Einstein
condensates with an aspect ratio of 152 using Bragg spectroscopy.
We observe the Lorentzian momentum distribution characteristic of
one-dimensional phase fluctuations. The temperature dependence of
the width of this distribution provides a quantitative test of
quasi-condensate theory. In addition, we observe a condensate
length consistent with the absence of density fluctuations, even
when phase fluctuations are large.
\end{abstract}
\maketitle
%
%%%%%%%%%%%%%%%%%%%%%%%%%%%%%%%%% body %%%%%%%%%%%%%%%%%%%%%%%%%%%%%%%%%
%
One of the most striking features of Bose-Einstein condensates is
their phase coherence. Extensive experimental work on dilute
atomic gases has demonstrated a uniform phase of three-dimensional
(3D) trapped condensates \cite{Phillips_coherence1,stenger99},
even at finite temperature \cite{Bloch}. In low dimensional
systems, however, phase fluctuations of the order parameter are
expected to destroy off-diagonal long range order (see
\cite{petrov1d, Stoof_Griffin_Castin} and references therein).
This phenomenon also occurs in sufficiently anisotropic 3D
samples, where phase coherence across the axis (long dimension) is
established only below a temperature $T_{\rm{\phi}}$, that can be
much lower than the critical temperature $T_{\rm{c}}$
\cite{petrov3d}. In the range $T_\phi < T < T_{\rm c}$, the cloud
is a ``quasi-condensate'', whose incomplete phase coherence is due
to thermal excitations of 1D axial modes, with wavelengths larger
than its radial size. Quasi-condensates in elongated traps have
been observed by Dettmer {\em et al.} \cite{dettmer01}, who
measured the conversion, during free expansion, of the phase
fluctuations into ripples in the density profile. Although the
conversion dynamics is well understood \cite{Ertmer3}, the
measured amplitude of density ripples was a factor of two smaller
than expected.

In this Letter, we report on the measurement of the axial
coherence properties of quasi-condensates via momentum Bragg
spectroscopy. In previous work using this technique
\cite{stenger99,davidson}, the finite size and mean-field energy
were the primary contributors to the spectral width. By contrast,
the dominant broadening in our conditions results from thermally
driven fluctuations of the phase, which reduce the coherence
length \cite{petrov1d,petrov3d}. Indeed, the axial momentum
distribution is the Fourier transform of the spatial correlation
function $C(s) = \int d^3 \mathbf{r}\,\langle
\hat{\Psi}^{\dagger}(\mathbf{r}-s\,\mathbf{u}_{\rm z}/2)
\hat{\Psi}(\mathbf{r}+s\,\mathbf{u}_{\rm z}/2)\rangle$
\cite{Zambelli}, where $\mathbf{u}_{\rm z}$ is the axial unit
vector. When phase fluctuations dominate (i.e. $T \gg T_\phi$),
the axial momentum width is hence proportional to $\hbar/L_\phi$,
where $L_\phi$ is the characteristic axial decay length of $C(s)$.
Experimentally, for $6<T/T_{\rm{\phi}}<36$, we find momentum
distributions with Lorentzian shapes, whose widths increase with
$T$. Such a shape is characteristic of large phase fluctuations in
1D \cite{Fabrice_theory}, which result in a nearly exponential
decay of $C(s)$. Moreover, the momentum width agrees
quantitatively with theoretical predictions to within our $15\%$
experimental uncertainty. This implies, in this temperature range,
a coherence length substantially smaller than the quasi-condensate
length $2L$, from about $L/18$ to $L/4$. We have also checked an
important feature of quasi-condensates: the suppression of density
fluctuations even in the presence of large phase fluctuations.

We produce a Bose-Einstein condensate of $^{87}$Rb in the
$5S\/_{1/2}$ $|F\!=\!1, \ m_F\!=\!-1\rangle$ state. A new design
of our iron-core electromagnet, with respect to our previous work
\cite{interrupted}, allows us to lower the bias field of the
Ioffe-Pritchard trap to obtain tighter radial confinement. Final
radial and axial trap frequencies are respectively
$\omega_\perp/2\pi=760(20)\,$Hz and $\omega_z/2\pi=5.00(5)\,$Hz.
The condensates, containing around $5 \times 10^4$ atoms
\cite{calib_N0}, are needle-shaped, with a typical half-length
$L\simeq130\,\mu$m and radius $R\simeq0.8\,\mu$m. The chemical
potential being a few times $\hbar \omega_\perp$, the clouds are
between the 3D and 1D Thomas-Fermi (TF) regime \cite{1D}. However,
the low-lying excitations of the condensate are 1D in character,
due to the large aspect ratio of the trap \cite{1D_Stringari}.

Our momentum distribution measurement is based on four-photon
velocity-selective Bragg diffraction. Atoms are diffracted out of
the condensate by interaction with a moving standing wave, formed
by two counter-propagating laser beams with a relative detuning
$\delta$ \cite{stenger99,davidson}. Due to the Doppler effect, the
momentum component resonantly diffracted out of the condensate is
$p_z=M(\delta-8\omega_{\rm{R}})/(2k_{\rm L})$ with
$\omega_{\rm{R}}=\hbar k_{\rm{L}}^2/(2M)$, $M$ the atomic mass,
and $k_{\rm L}=2\pi/\lambda$ ($\lambda=780.02$\,nm). The lasers
are tuned 6.6\,GHz below resonance to avoid Rayleigh scattering.
The laser intensities (about 2 mW/cm$^2$) are adjusted to keep the
diffraction efficiency below 20\,\%.

To build the momentum spectrum of the quasi-condensate, we measure
the fraction of diffracted atoms versus the detuning $\delta$
between the counter-propagating laser beams. The differential
frequency $\delta$ must be stable to better than the desired
spectral resolution, about 200\,Hz for our typical
$L_\phi=10\,\mu$m. The optical setup is as follows. A single laser
beam is spatially filtered by a fiber optic, separated into two
arms with orthogonal polarizations, frequency shifted by two
independent 80\,MHz acousto-optic modulators, and recombined. The
modulators are driven by two synthesizers stable to better than
1\,Hz over the typical acquisition time of a spectrum. The
overlapping, recombined beams are then sent through the vacuum
cell, parallel (to within 1\,mrad) to the long axis of the trap,
and retro-reflected to obtain two standing waves with orthogonal
polarizations, moving in opposite directions. After we mounted the
critical retro-reflecting mirror on a long, rigid plate to
minimize axial vibrations, the average over ten beat notes had a
half-width at half-maximum (HWHM) of $216(10)$\,Hz for a 2-ms
pulse \cite{beat}.

\begin{figure}[!t]
\includegraphics[width=8.5cm]{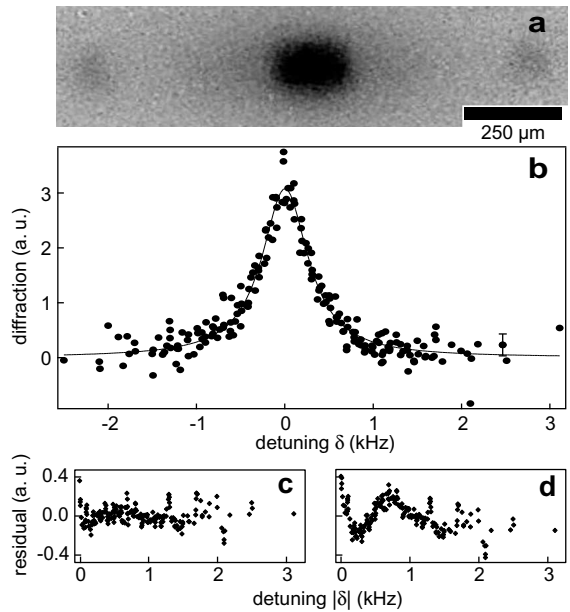}
\caption{{\bf (a)} Absorption image of degenerate cloud (center)
and diffracted atoms (left and right), averaged over several
shots, after free-flight expansion. {\bf (b)} Diffraction
efficiency (rescaled) versus relative detuning of the Bragg lasers
at $T=261(13)$\,nK, corresponding to $T/T_\phi=20(2)$. A typical
statistical error bar is shown. This spectrum is the superposition
of 12 ``elementary spectra'', as described in the text. The true
average center is 30.18(2)\,kHz, close to 30.17\,kHz, the
four-photon recoil frequency. The solid line is a Lorentzian fit,
giving a half-width at half-maximum (HWHM) of 316(10)\,Hz. {\bf
(c)} and {\bf (d)} show respectively the residual of a Lorentzian
and of a Gaussian fit to the above spectrum. Residuals are folded
around the $\delta=0$ axis, and smoothed with a six-point-wide
sliding average. } \label{spectres}
\end{figure}

The following experimental procedure is used to acquire a momentum
spectrum. At the end of forced evaporative cooling, the radio
frequency knife is held fixed for 6.5\,s to allow the cloud to
relax to equilibrium. Indeed, we observe axial shape oscillations
of the cloud, triggered by the onset of Bose-Einstein condensation
\cite{shvarchuck02,Fabrice_proceed}, despite a slow evaporation
(less than 100\,kHz$/$s) across $T_{\rm c}$. The magnetic trap is
then switched off abruptly, in roughly 100\,$\mu$s, and the cloud
expands for 2\,ms before the Bragg lasers are pulsed on for 2\,ms.
Since there are two laser standing waves moving in opposite
directions, two diffracted components are extracted. We wait for a
further 20\,ms to let the diffracted atoms separate from the
parent condensate, and take an absorption image
(Fig.~\ref{spectres}a). Diffraction efficiency is defined as the
fraction of atoms in each secondary cloud. We repeat this complete
sequence for several detunings (typically 15), several times
(typically 5). After averaging the diffraction efficiencies
measured at each detuning $\delta$, we obtain two ``elementary
spectra'', one for each diffraction component.

We take the Bragg spectrum after expansion rather than in the trap
to overcome two severe problems. In the trapped condensate, first,
the mean free path (about 10\,$\mu$m) is much smaller than its
axial size, typically 260\,$\mu$m, so that fast Bragg-diffracted
atoms would scatter against the cloud at rest
\cite{Chikkatur_collisions}. Second, the inhomogeneous mean field
broadening \cite{stenger99} would be of the order of 300\,Hz, i.e.
larger than the spectral width expected from phase fluctuations.
By contrast, after 2\,ms of free expansion, the peak density has
dropped by two orders of magnitude \cite{castin_kagan}, and both
effects become negligible. The measured momentum spectra
faithfully reflect the phase fluctuations of the trapped
quasi-condensate. Indeed, the phase fluctuations do not
significantly evolve in 2\,ms, since the typical timescale for
their complete conversion into density ripples varies from 400\,ms
to 15\,s for the range of temperatures we explore
\cite{Fabrice_theory}. Also, the mean field energy is released
almost entirely in the radial direction, because of the large
aspect ratio of the trap \cite{castin_kagan}, and contributes only
about $50$\,Hz of Doppler broadening in the axial direction. The
only perturbation due to the trap release seems to be small axial
velocity shifts (around $100\,\mu$m/s) attributed to stray
magnetic gradients that merely displace the spectra centers.

Bragg spectra have been taken at various temperatures between
90(10)\,nK and 350(20)\,nK, while $T_{\rm c}$ varied from
280(15)\,nK to 380(20)\,nK. The temperature was fixed to within
20\,nK by controlling the final trap depth to a precision of
2\,kHz, and measured from a fit to the wings of an averaged
absorption image. The fitting function is an ideal Bose
distribution with zero chemical potential, plus an inverted
parabolic profile for the quasi-condensed cloud. At each
temperature, pairs of elementary spectra (described above) were
collected across a 125-ms-wide range of hold times to average over
residual oscillation and slowly varying fluctuations. All
elementary spectra corresponding to the same temperature are
reduced to the same surface, background, and center, and
superposed (Fig.~\ref{spectres} b).

The line shape of the resulting spectra is clearly closer to a
Lorentzian than to a Gaussian (see Fig.~\ref{spectres}). This is a
significant result, because a Lorentzian-like profile is expected
for a momentum distribution dominated by phase fluctuations (see
\cite{Fabrice_theory} and below), in contrast to the gaussian-like
profile expected for a pure condensate \cite{Zambelli,stenger99}.
From the Lorentzian fit, we extract the measured half-width
$\Delta\nu_\textrm{M}$ for each temperature.

The theoretical results obtained in \cite{petrov3d,
Fabrice_theory} do not apply directly to our experiment. In those
works, a 3D Thomas-Fermi density profile was assumed, whereas in
our case modification of the density profile by the thermal cloud
and radial quantum pressure must be accounted for. We find (and
discuss below) that a parabolic profile is still a good fit
function, but that the usual $T=0$ relations between $\mu$, $L$,
$R$, and the number of condensed atoms $N_0$ are no longer valid.
We therefore extend the calculation of the axial correlation
function in \cite{Fabrice_theory} to an arbitrary density profile,
in the local density approximation. In the mean field regime
\cite{petrov1d}, from the result of \cite{kane_kadanoff} for a 1D
uniform Bose gas at finite temperature, we obtain: {\small
\begin{equation}C(s)=\int dz\, n_{1}(z)\,{\rm
exp}(-\frac{n_{1}(0)|s|}{2n_{1}(z)L_\phi}),\label{equ:C}
\end{equation}}where $n_{1}(z)=\int d^2{{\bf r}_\perp}\,n_0({\bf r}_\perp,z)$ is
the axial 1D density of the quasi-condensate, while $n_0({\bf r})$
is its 3D density profile. The coherence length near the center of
the trap is given by $L_\phi=\hbar^2 n_{1}(0)/(Mk_{\rm B}T)$
\cite{TFlimits}. Following Petrov {\it et al.}
\cite{petrov1d,petrov3d}, we define the temperature which
delineates the border between coherent and phase-fluctuating
condensates as $T_{\phi}=L_\phi T /L$. Since $n_1(0)$, $L$, and
$T$ are extracted directly from the images, the definitions of
$L_\phi$ and $T_\phi$ relate the coherence properties to
experimentally measured quantities. The axial momentum
distribution follows from a Fourier transform of $C(s)$ and is
well approximated by a Lorentzian of width $\Delta p_{\phi} =
\alpha \hbar/L_\phi$ (HWHM), with $\alpha=0.67$ for a parabolic
$n_ 0({\bf r})$ \cite{Fabrice_theory,alpha}. The predicted
spectral width is therefore $\alpha \Delta\nu_\phi$, where {\small
\begin{equation} \Delta \nu_{\phi}= \frac{2\hbar k_{\rm L}} {2 \pi
M L_{\phi}}.
\end{equation}}
In the following we will use $\alpha$ as a free parameter to test
the theory outlined above.

\begin{figure}[t!]
\includegraphics[width=8.5cm]{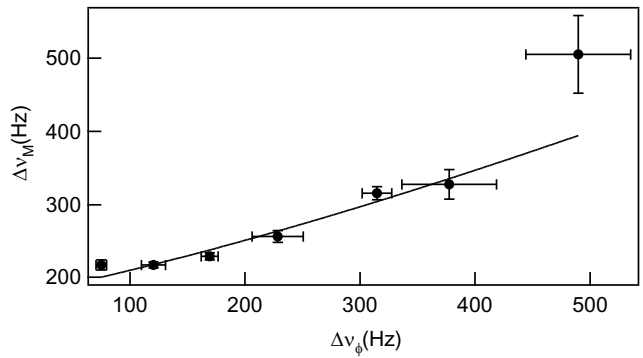}
\caption{Half-widths at half-maximum $\Delta\nu_{\rm M}$ of the
experimental Bragg spectra versus the parameter
$\Delta\nu_\phi\propto \hbar / L_\phi$ (see text). Vertical error
bars are the standard deviations of the fit width; the horizontal
error bars are the one-sigma statistical dispersions of
$\Delta\nu_\phi$. The solid line is a fit assuming a Voigt profile
for the spectra.} \label{largeur_vs_temp}
\end{figure}

Figure~\ref{largeur_vs_temp}
shows the measured spectral width $\Delta\nu_\textrm{M}$ versus
$\Delta \nu_\phi$. Dispersion of the measured $n_1(0)$ and $T$
results in the horizontal error bars on $\Delta\nu_\phi$, while
vertical error bars indicate the standard deviation of the fit
width. The measured widths increase at higher $\Delta \nu_\phi$ as
expected. To compare these data to theory, we need to take into
account a finite ``instrumental'' width of the Bragg spectra,
including the effect of the mirror vibrations, residual sloshing
in the trap, and Fourier broadening due to the 2-ms pulse length
(125\,Hz HWHM). We assume that all experimental broadenings result
in a Gaussian apparatus function of half-width $w_{\rm G}$, to be
convolved by the Lorentzian momentum profile with a half-width
$\alpha\Delta\nu_\phi$. The convolution, a Voigt profile, has a
half-width $\alpha\Delta\nu_\phi/2+\sqrt{w_{\rm
G}^2+(\alpha\Delta\nu_\phi)^2/4}$. Note that fitting a Voigt
profile instead of a Lorentzian to a spectrum gives the same total
HWHM to less than 5\,\%, but the Lorentzian shape is too
predominant to extract reliably the Gaussian and the Lorentzian
contributions to the profile. Using $\alpha$ and $w_{\rm G}$ as
free parameters to fit the data of Fig.~\ref{largeur_vs_temp}, we
find $w_{\rm G}=176(6)$\,Hz, and $\alpha= 0.64(5)(5)$. The first
uncertainty quoted for $\alpha$ is the standard deviation of the
fit value. The second results from calibration uncertainties on
the magnification of the imaging system and on the total atom
number, which do not affect $w_{\rm G}$. The agreement of the
measured value of $\alpha$ with the theoretical value 0.67, to
within the 15\,\% experimental uncertainty, confirms
quantitatively the temperature dependence of the momentum width
predicted in Ref.~\cite{petrov3d}. The coherence length
$\hbar/p_\phi$ deduced from this measurement varies between
$5.9(8)$ and $39(4)\,\mu$m, in the range $6<T/T_\phi<36$.

Another important aspect of quasi-condensates, the suppression of
axial density fluctuations, is investigated here through the size
of the trapped condensate. In the presence of density
fluctuations, the resulting interaction energy \cite{ketterle_g2}
would increase the size with respect to the expectation for a
smooth density profile. The measured axial half-length $L$ after
release faithfully reflects the half-length in the trap since
axial expansion is negligible \cite{expansion}.
\begin{figure}[!t]
\includegraphics[width=8.5cm]{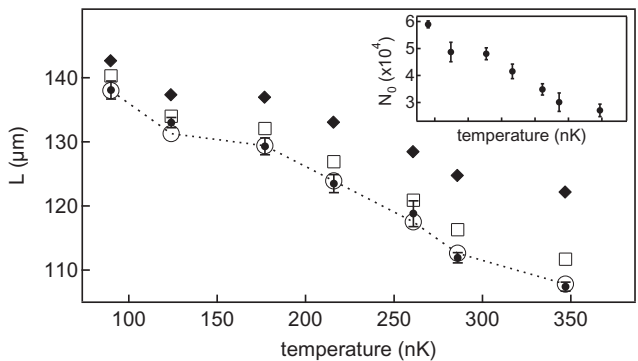}
\caption{Half-length $L$ of the quasi-condensate versus
temperature. Experimental values ($\bullet$ with statistical error
bars) are compared to three calculations: standard $T=0$
Thomas-Fermi ($\blacklozenge$) and Eq.~\ref{eq:L} in the text,
with and without radial quantum pressure ($\cdots\cdots$\hspace
{-17.8pt}$\bigcirc$\hspace {+7pt} and $\square$ respectively).
Calculations use the measured temperature and number of condensed
atoms, as determined by a two-component fit to averaged absorption
images. {\bf Inset}: Number of condensed atoms $N_0$ with
statistical dispersion at each temperature.}
\label{longueur_vs_tsurtphi}
\end{figure}
Figure~\ref{longueur_vs_tsurtphi} shows that the measured values
of $L$ at various temperatures (filled circles) are smaller than
the standard Thomas-Fermi prediction (diamonds) $L_{\rm
TF}^2=2\mu/(M\omega_z^2)$, with $\mu$ given by
$2\mu=\hbar\bar{\omega}(15N_0a/\sigma)^{2/5}$ \cite{DalfovoRMP},
where $\bar{\omega}=(\omega_\perp^2\omega_z)^{1/3}$,
$\sigma=\sqrt{\hbar/(M\bar{\omega})}$ and $a=5.32$\,nm
\cite{scat_length}. We find that this smaller value results from
the compression of the quasi-condensate by the 3D thermal
component (excitations with energy much larger than
$\hbar\omega_\perp$), and from radial quantum pressure. The 3D
excited states contribute negligibly to the fluctuations of the
phase \cite{petrov3d} and only the density profile is affected.
Using a Hartree-Fock approach \cite{intmodel,DalfovoRMP}, we find:
{\small\begin{equation}
 L^2=\frac{2g}{M\omega_z^2}\left\{n_0(0)+\frac{2}{\lambda_{\rm T}^{3}}
 \left[{\textsl g}_{3/2}(e^{-\frac{gn_0(0)}{k_{\rm B}T}})-{\textsl
 g}_{3/2}\left( 1 \right)\right]\right\},
\label{eq:L}\end{equation}}with the coupling constant
$g=4\pi\hbar^2a/M$, the thermal de Broglie wavelength
$\lambda_{\rm T}=[2\pi\hbar^2/(M k_{\rm B} T)]^{1/2}$, and
${\textsl g}_{3/2}(x)=\sum _{n=1}^{\infty}x^n/n^{3/2}$. The open
circles in Fig.~\ref{longueur_vs_tsurtphi} show the solution of
Eq.~(\ref{eq:L}) assuming a parabolic profile, such that
$n_0(0)=15N_0 L^{-3}\epsilon^2/(8\pi)$. The aspect ratio
$\epsilon$ is calculated according to the theory developed in
\cite{Zubarev_1D}, which takes into account radial quantum
pressure. The calculated lengths are in agreement with our
measurements to better than both the statistical error (shown),
and our estimated calibration uncertainty (not shown) of 4\,\% on
the ratio $L_{\rm meas.}/L_{\rm calc.}$. We conclude that a
phase-fluctuating condensate has the same smooth profile as a true
condensate, and thus that the axial density fluctuations are
suppressed even when phase fluctuations are large.

In conclusion, we have demonstrated three important features of
quasi Bose-Einstein condensates: (i) the momentum distribution
shape, found Lorentzian; (ii) the temperature dependence of the
momentum width; and (iii) the suppression of density fluctuations.
Our results are in quantitative agreement with the
finite-temperature, interactive theory developed in
\cite{petrov1d,petrov3d} supplemented by a Hartree-Fock treatment
of 3D excited states. The same method could be applied to
investigate how long range order develops during the condensate
growth (see \cite{shvarchuck02} and references therein).

\begin{acknowledgments}
We thank M.\ L\'ecrivain and V.\ Boyer for the development of the
electromagnet trap used in this work, and J. Retter for useful
comments on the manuscript. We also thank G.\ V.\ Shlyapnikov and
D.\ S.\ Petrov for stimulating interactions, and S.\ Gupta and N.\
Davidson for useful conversations. This work was supported by the
CNRS, the DGA, and the EU.
\end{acknowledgments}

% Create the reference section using BibTeX:

%
%

\begin{thebibliography}{99}
\bibitem{calib_N0}{The atom number measurement is calibrated to a 20\,\% precision
by comparing a measurement of $N_{\rm c}$ to theory
\cite{DalfovoRMP}.}
\bibitem{TFlimits}{Using the relevant expressions for $n_{1}(0)$, the results obtained in \cite{petrov1d} for the 1D case and
\cite{petrov3d} for the 3D case are recovered, in the limit
$L_{\phi} \ll L$.}
\bibitem{expansion}{We have measured the axial
size of quasi-condensates after various free flight times (less
than 25\,ms). No measurable expansion was found.}
\end{thebibliography}
\end{document}